\documentclass[amsmath,amssymb,aps,prx,preprint]{revtex4-1}
\pdfoutput=1 

\usepackage[english]{babel}
\usepackage{graphicx}
\usepackage{dcolumn}
\usepackage{bm}
\usepackage{subfloat}
\usepackage{subfig}
\usepackage{color}

\begin{document}

\title{Local Dimension of Complex Networks}

\author{Filipi Nascimento Silva}
\email{filipinascimento@gmail.com.}
\author{Luciano da Fontoura Costa}
\email{ldfcosta@gmail.com.}
\affiliation{Institute of Physics at S\~ao Carlos - University of S\~ao Paulo \\ Luciano Costa's Group}
\homepage[]{http://cyvision.ifsc.usp.br/cyvision/}

\date{August 11, 2013}
\begin{abstract}
Dimensionality is one of the most important properties of complex physical systems.  However, only recently this concept has been considered in the context of complex networks. In this paper we further develop the previously introduced definitions of dimension in complex networks by presenting a new method to characterize the dimensionality of individual nodes. The methodology consists in obtaining patterns of dimensionality at different scales for each node, which can be used to detect regions with distinct dimensional structures as well as borders. We also apply this technique to power grid networks, showing, quantitatively, that the continental European power grid is substantially more planar than the network covering the western states of US, which present topological dimension higher than their intrinsic embedding space dimension. Local dimension also successfully revealed how distinct regions of network topologies spreads along the degrees of freedom when it is embedded in a metric space.

\end{abstract}
\maketitle

\emph{Dimension} is one of the most basic concepts in Physics. Diffusion\cite{diffusionWeiss1994,diffusionHavlin2005}, waves propagation\cite{hdobbins:Jackson1998Classical}, Brownian motion\cite{Nelson98dynamicaltheories}, as well as many other physical processes are highly influenced by the dimension in which those phenomena take place. Dimensionality also allows us to quantify the degrees of freedom in a system, as well to characterize the macroscopic dynamics on complex systems by means of statistical mechanics\cite{Huang:1987fk}. Representations of complex systems by complex networks\cite{Costa:2011fk, Barabasi:1999p279, Newman:2003p274} has proven to be generally successful to describe their various features without losing their intrinsic complexity. Important physical dynamics, like diffusion and information propagation can take place in such structures. Surprisingly, not much attention has been given to characterizing the dimensionality of complex networks. 

Some early attempts on describing the dimensionality of networks\cite{ERDOS:1965ij,Harary:1976,MAULDIN1986} took into account well-known regular lattices and small graphs. Characterization of the dimensionality of complex networks was first introduced by Cs\'anyi\cite{Csanyi2004}, and was further developed by Gastner and Newman\cite{Gastner2006}. They presented a flexible way to calculate the dimension of arbitrary networks in terms of the scaling property of the topological volume. In another work, Shanker\cite{Shanker2007} generalized this concept by developing a new and mathematically coherent definition of global dimension for complex networks based on the Riemann zeta function for graphs. He also proves that unbounded random and small-world networks present infinite dimension\cite{Shanker2008}.

Recently, Daqing et al \cite{Havlin2011} introduced three novel methods to obtain the dimensionality of networks, which yield the same results even for distinct dynamics such as diffusion, random walks and percolation. They found that the dimension values depend neither on the size nor on the average degree of networks.  The dimension measurements proposed recently\cite{Gastner2006,Havlin2011}, provide good insights about the global dimensional structure of networks, but cannot characterize the nodes individually.

In his paper, Gastner commented that the dimensionality may change considerably among the vertices of a network, therefore a local dimension measurement should be necessary to better characterize such systems, however no further works explored the dimensional features of individual vertices. Interdependent networks, for example, may encompass networks with distinct dimensions. In this case the measurement of global dimension does not represent all nodes in the network.

In this paper we further investigate the use of dimensionality measurements to characterize two power grid networks, namely: the continental european network (\emph{EU power grid}) and the western states power grid of the United States (\emph{US power grid}). To characterize the dimensionality of individual nodes in a network, we propose a extension for the measurement of the dimension, which takes into account the local scaling property of the volume at different topological distances from a node. We extensively apply the new measurement to geographical and small-world models by taking into account the changes of the dimensionality for various values of rewiring probability and average node degree. Also, we compare the new measurement with the \emph{accessibility}\cite{Travencolo2008Accessibility, Travencolo2009Border, Costa2010Transportation} which also quantifies the degrees of freedom for nodes in the sense of the effective number of accessed nodes for given topological distance. We also illustrate the application of this methodology to understanding how is the redistribution of nodes when a network is embedded in a higher dimensional space.

The dimension of a network can be understood as the minimum $d_e$ for which the entire network can be embedded in a $d_e$-dimensional space without losing its topological structure. This concept was early introduced by Erd\"os\cite{ERDOS:1965ij} but the presented methodology was shown to be a NP-Complete\cite{Dailey1980289} problem, which turns out to be unfeasible for determining the dimensionality of large complex networks.

Most of the space embedded real networks that also are not small-world present length distributions that follows a power-law\cite{Lambiotte20085317,havlin2008}. This means that for each concentric ball\cite{Costa:2006p278,Costa:2008p14}, $B_i(r)$, centralized on a node $i$ the number of nodes within obeys the relationship $B_i(r) \sim r^d$ with regard to the topological distance $r$.

The constant $d$ is called dimension because on a infinite $n$-dimensional lattice, $d$ correspond exactly to $n$. In fact, the dimension $d$ characterizes the diffusion processes on networks in the same way as spatial dimension characterizes the diffusion on regular spaces. For example, a diffusion process occurring on a network with $d=2$ will present the same pattern of diffusion as if it was occurring on a 2D-plane. An estimate value of the dimension of networks can be obtained by linear regression of $B_i(r)$ on the log-log scale considering a sample of central nodes\cite{Havlin2011}.

\begin{figure*}[!htbp]
 \centering
 \subfloat[]{\label{fig:growBall}\includegraphics[height=6.5cm]{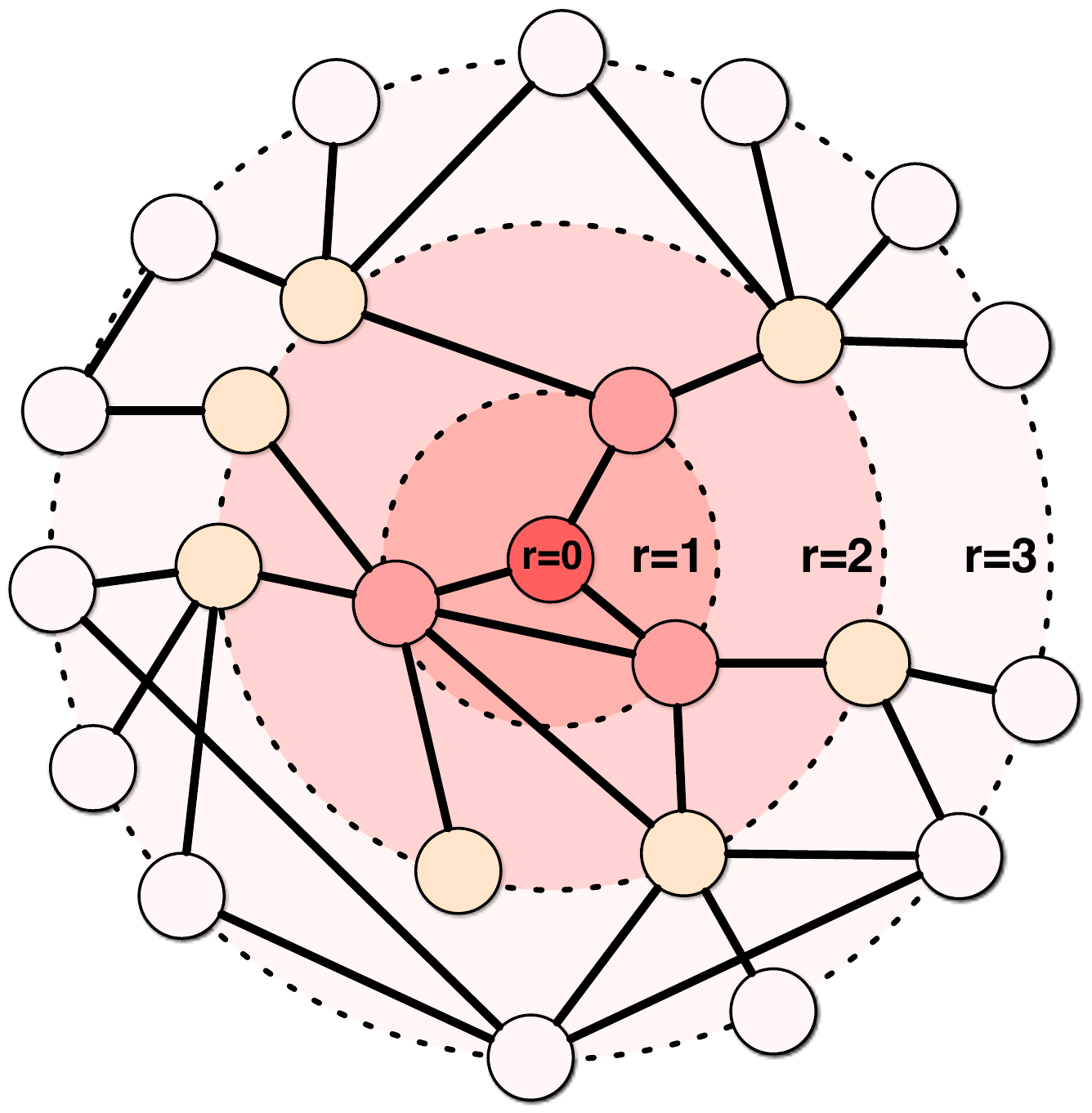}}
~
 \subfloat[]{\label{fig:growDistribution}\includegraphics[height=6.0cm]{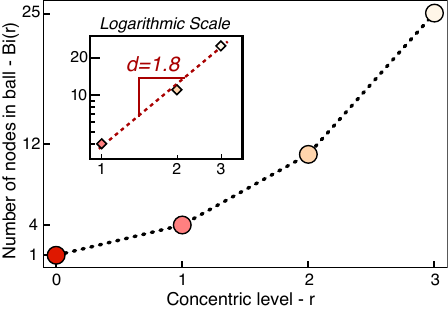}}
 \caption{Example of a network represented by concentric balls, depicted as filled circles in ({\bf a}),  starting from a central node up to the level $r=3$. The number of nodes on each ball is presented in ({\bf b}), the angular coefficient of the double logarithmic curve yields the dimension coefficient, shown in the inset. }
 \label{fig:concentric}
\end{figure*}

While the concept of global dimension provides important characteristics about the embedding space and dynamical process on networks, substantially richer information can be obtained by characterizing not only the dimension of the entire network but also the \emph{local dimension of nodes}.

\begin{figure*}[!htbp]
 \centering
 \subfloat[Local dimension for $r=2$.]{\label{fig:cityExampleD3}\includegraphics[width=5.5cm]{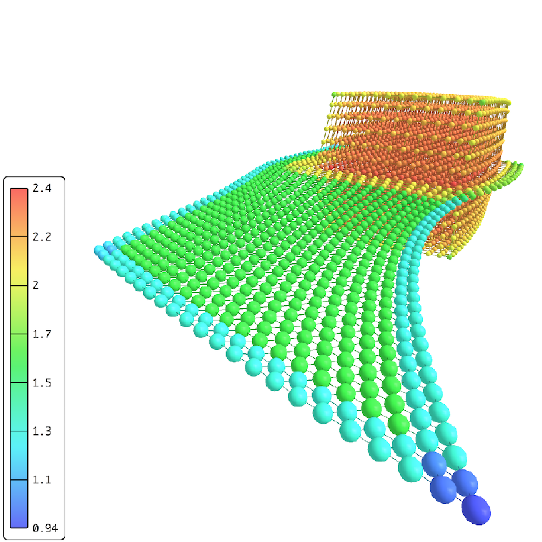}}
~
 \subfloat[Local dimension for $r=4$.]{\label{fig:cityExampleD4}\includegraphics[width=5.5cm]{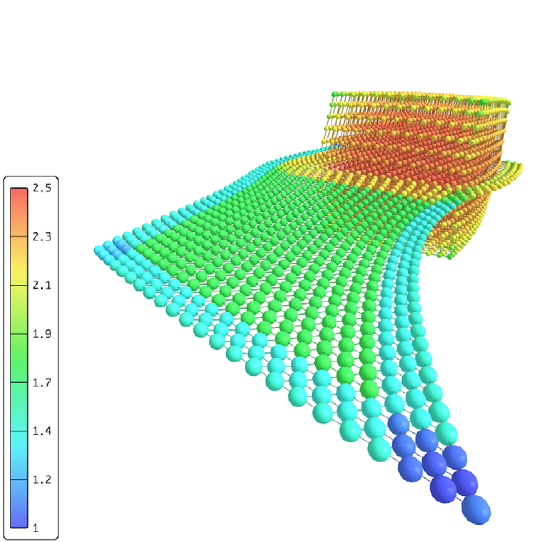}}
 \caption{Example of mixed regular network of a 2D and 3D lattices, generated in a way that node degree is constant along all its topology. Two levels of local dimension are shown (represented by a spectral color scale), both highlight the contrast between the two distinct dimensional regions on the network. Because of the discrete nature of networks, the dimension scale may be lower than the expected dimension of planar and 3D embedding spaces.}
 \label{fig:cityExample}
\end{figure*}

Because of the heterogeneous topologies found in most real world networks, the distribution of nodes $B_i(r)$ along the concentric distance $r$ may vary greatly for individual nodes as starting points~\cite{Costa:2008p14} and may not follow a strict power law. Those networks display multi-dimensional structure, therefore it is not possible to assign a unique dimension value for the entire topology, for example, in the outskirts of a city, most of the social interactions are embedded on a 2D-plane, while  downtown the social interactions may take place on larger, multiple floor buildings (Figure~\ref{fig:cityExample}).

We can further develop the ball growth relationship $B_i(r) \sim r^d$, by considering that the dimension, $D_i(r)$, may vary both for each starting node $i$ as well for each concentric level $r$. The dimension coefficient can be obtained by the slope of the $B_i(r)$ curve on a double logarithmic scale, as follows:

\begin{align}
 B_i(r) &= \alpha \, r^{D_i(r)}\\
 \label{eq:dimensionderivative}
 D_i(r) &= {d \over d\log r}\log B_i(r)
\end{align}

This definition of dimensionality is closely related to the multi-scale fractal dimension\cite{Backes201044} obtained by using the Bouligand-Minkowsk method\cite{Backes2008} on 2D shapes, however in this case we are measuring the local dimensional limit of the space around a point (node) (i.e. not measuring fractal dimension of any special shape or set of points).

The derivative in equation \ref{eq:dimensionderivative} can be expressed in terms of $r$ and $B_i(r)$, and because of the discrete nature of such measurements, it can also be discretized as follows:

\begin{align}
 D_i(r) & = {r \over B_i(r)}{d \over d r}B_i(r) \\
 \label{eq:dimensiondiscrete}
 D_i(r) & \simeq r {n_i(r) \over B_i(r)}
\end{align}

where $n_i(r)$ is the number of nodes on the ring at concentric level $r$, i.e. the number of nodes distancing exactly $r$ from the central node.

Even considering the approximation on equation~\ref{eq:dimensiondiscrete}, which took into account the discrete nature of complex networks, it is important to note that the equation is still valid in the case of continuous surfaces. Considering the special case of the continuous 2D plane, where $n(r) = 2 \pi r$ and $B(r) = \pi r^2$, the equation yields the expected result $D=2$.

An example of generated multi-dimensional network is shown in figure~\ref{fig:cityExample}, alongside with the calculations of local dimension $D_i(r)$ for $r=2$ and $r=4$.  Taking distances from starting nodes up to the third concentric level, i.e. $r=3$, the measurement of local dimension, $D_i(3)$, sets apart the two distinctive dimensional regions, even considering nodes with the same degree. It is important to note that because of the finite size of the network, nodes near the frontier tends to present lower local dimensionality.

\begin{figure*}[!htbp]
 \centering
  \subfloat[Local dimension.]{\label{fig:geographicNetwork}\includegraphics[height=5.48cm]{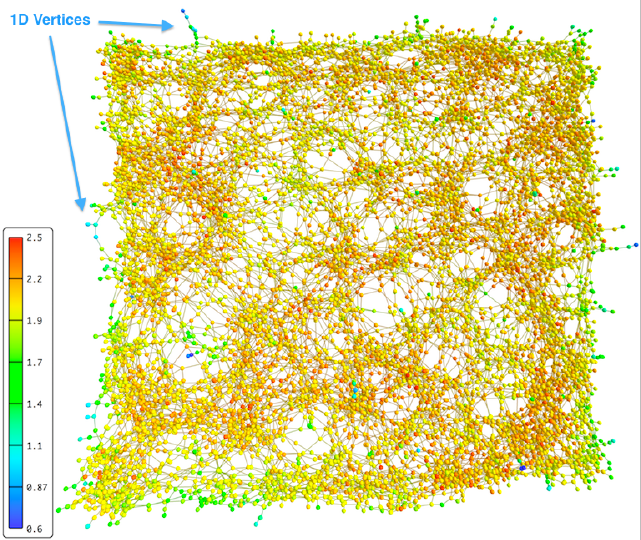}}
  ~
  \subfloat[Average local dimension along $r$.]{\label{fig:geoAverageDimension}\includegraphics[height=5.2cm]{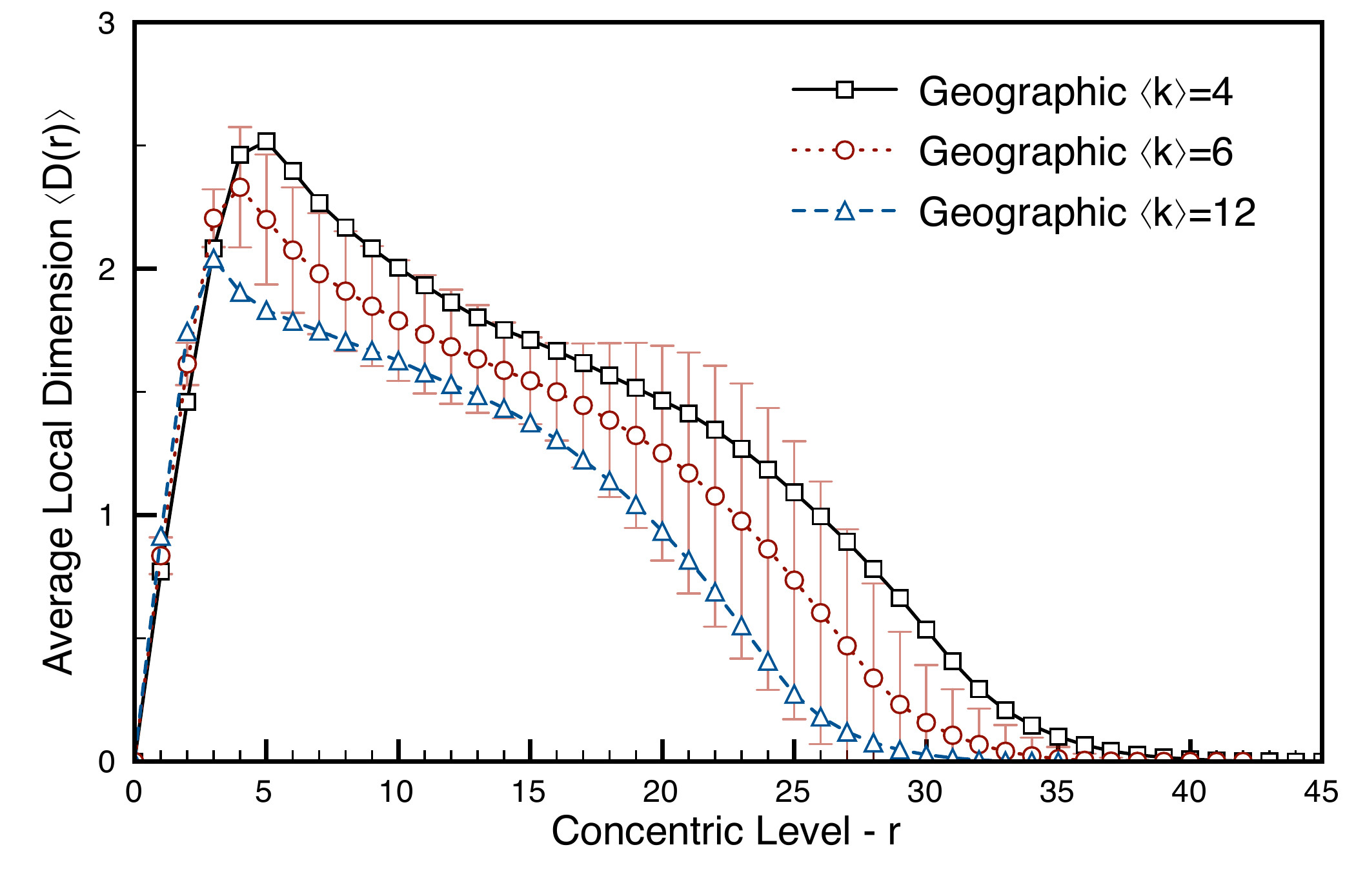}}
 \caption{The local dimension ($r=4$) for a geographic model with $\langle k \rangle = 4$ is shown in ({\bf a}), where red and orange regions present the highest values of local dimension while nodes drawn in blue and green present lower dimensionality ($d\approx1.0$). ({\bf b}) shows the average local dimension for each concentric distance, taken over all the nodes of three geographical models network with size $N=10000$ and distinct values of node degree, namely $\langle k \rangle = 4$, $\langle k \rangle = 6$ and $\langle k \rangle = 12$. For the sake of clearness only the standard deviation for the curve with $\langle k \rangle = 6$ is presented as error bars, the other curves displayed similar behavior.}
 \label{fig:exampleGeo}
\end{figure*}

By using equation~\ref{eq:dimensiondiscrete}, we obtained curves of local dimension $D_i(r)$ for every node on random 2D geographical network models\cite{Dall2002,Barthelemy20111} generated with different values of average degree $\langle k \rangle$. Figure~\ref{fig:geographicNetwork} shows the 2D projection of a geographical network model with $\langle k \rangle=4$ where colors represent the local dimension of each node at $r=4$. Clustered regions are prone to present high local dimension while regions of lower dimension  are pathways between such clusters or border regions where some 1D lines of nodes are presented (as indicated in the figure). Next, we calculated the average value of local dimension taken over the nodes,  $\langle D(r)\rangle$, resulting in curves of dimensionality patterns for each geographic network, which are shown in figure~\ref{fig:geoAverageDimension}. All curves are characterized by a peak followed by a mostly concave decay with their maximum value depending on $\langle k \rangle$, also their values exceed the expected embedding dimension 2. This effect is a consequence of the fact that random geographical networks are known to be highly clustered\cite{Dall2002} due to prohibitive probability of nodes presenting long-range connections, which in turn results in higher values of dimensionality for the first few neighborhoods of nodes. This effect is diminished for networks with large $\langle k \rangle$ because the network becomes much more similar to the 2D lattice with large neighborhood count. For distant concentric levels $r$, the dimensionality drops to null as there is only a small number of nodes after certain distances, this effect reflects the fact that a discrete and bounded space behave like a point, therefore the dimension for all nodes is $D_i(r \to \infty)=0$.

\begin{figure*}[!htbp]
 \centering
 \includegraphics[width=12.9cm]{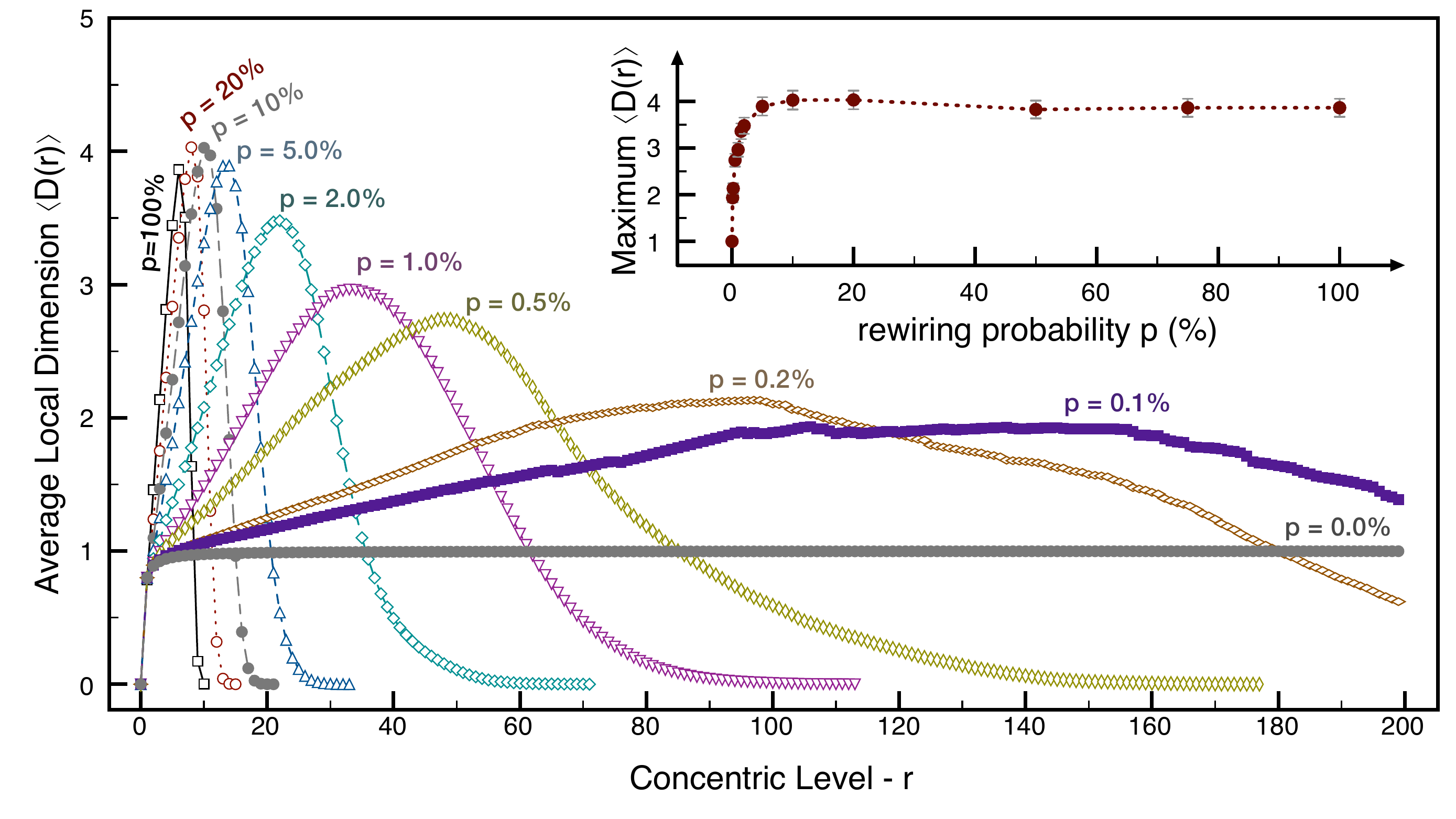}
 \caption{Average local dimension distribution of the Watts-Strogatz 1-dimensional network model for various configurations of rewiring probability $p$ with fixed $N=5000$ and $\langle k\rangle=4$. The inset displays the curve for the dependence of the maximum value of $\langle D(r)\rangle$ with $p$, which stabilizes for local dimension $\approx 4$.}
 \label{fig:smallworldDimension}
\end{figure*}

To better understand the effects of long-range connections for the local dimensionality we also obtained $\langle D(r)\rangle$ distributions for several small-world networks generated using the Watts-Strogatz model with different values for the rewiring probability $p$. Figure~\ref{fig:smallworldDimension} shows $\langle D(r)\rangle$ distributions for these networks. Similarly to the results obtained for geographical networks, the curves are characterized by a smooth peak with its height increasing with $p$. As expected, the diameter of the networks decreases with $p$, thus each curve becomes progressively wider by lowering $p$. For $p=0$ (i.e. the case when the network is a 1D lattice) the curve stabilizes for $\langle D(r)\rangle_{p=0}=0$, on the other hand, for $p=100\%$ (i.e. the case where the network is random) most of the nodes are accessed in the first $4$ concentric levels, yielding a very narrow peak. This is a consequence of the fact that the dimensionality of random networks is infinite, but because of the finite size of the networks the dimensionality scales with the number of nodes, thus limiting its maximum value. This effect is better displayed on the inset of figure~\ref{fig:smallworldDimension} which shows that the maximum local dimension escalates rapidly with $p$ until reaching $\max (\langle D(r)\rangle) \approx 4$ for $p>5\%$, this indicates that the networks began to present the dimensionality of random networks with only a few possible long-range connections(less than $5\%$ of edges).

Power grid networks are geographical structures already embedded in a 2D space, however they may feature a small number of long range connections, as well as heterogeneous node densities along the plane. These complex characteristics may give rise to a considerable difference between the embedding and topological dimensions on such networks.

The EU power grid\cite{rosas-casals:topological} network encompasses the entire continental european area considering the year 2003, covering most of the stations and power lines previously coordinated by the Union for the Coordination of Transmission of Electricity (UCTE), it contains $2783$ nodes and has $\langle k\rangle = 2.8$. The US power grid network\cite{Watts:1998kx} was taken from the Pajek dataset \cite{pajekdataset}, which covers the western states of United States, totalizing $4941$ nodes and with $\langle k\rangle = 5.3$. Both networks projections are shown in figure~\ref{fig:powergridProjections}.

\begin{figure*}[!htbp]
 \centering
 \subfloat[EU Power Grid.]{\label{fig:EUPowergrid2D}\includegraphics[height=5.5cm]{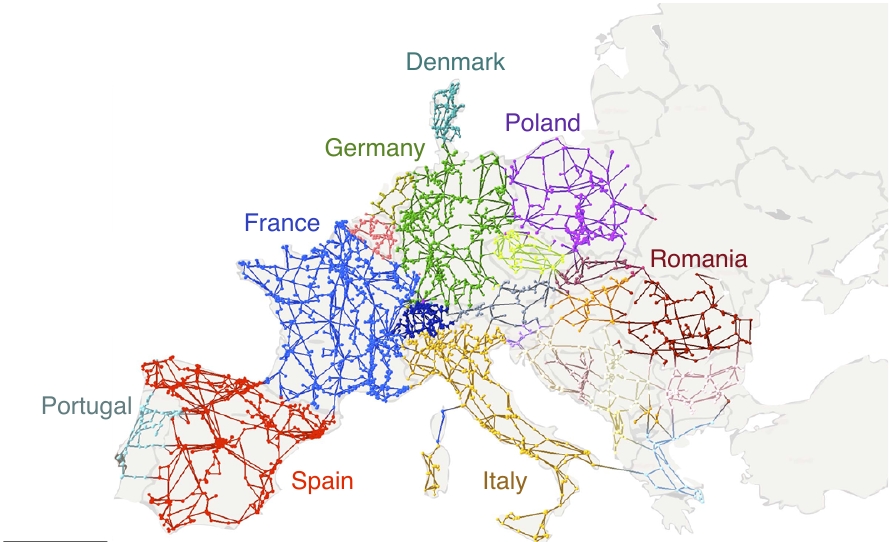}}
~
 \subfloat[US Power Grid.]{\label{fig:USPowergrid2D}\includegraphics[height=6.0cm]{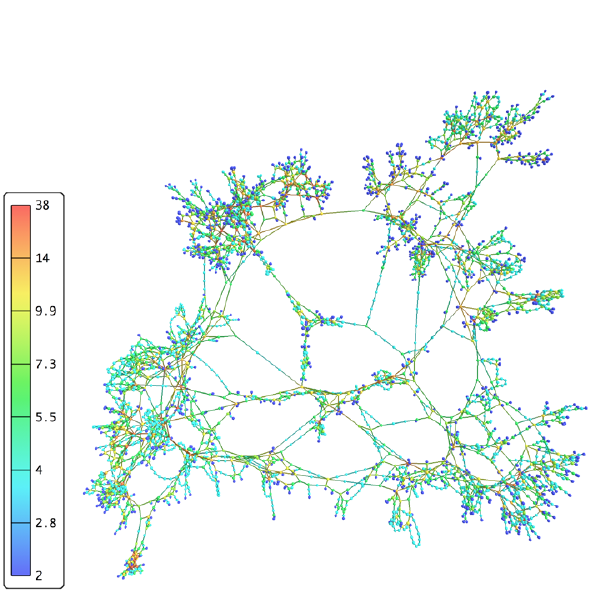}}
 \\
 \subfloat[Local Dimension.]{\label{fig:averageDimensionVsLevel}\includegraphics[height=6.0cm]{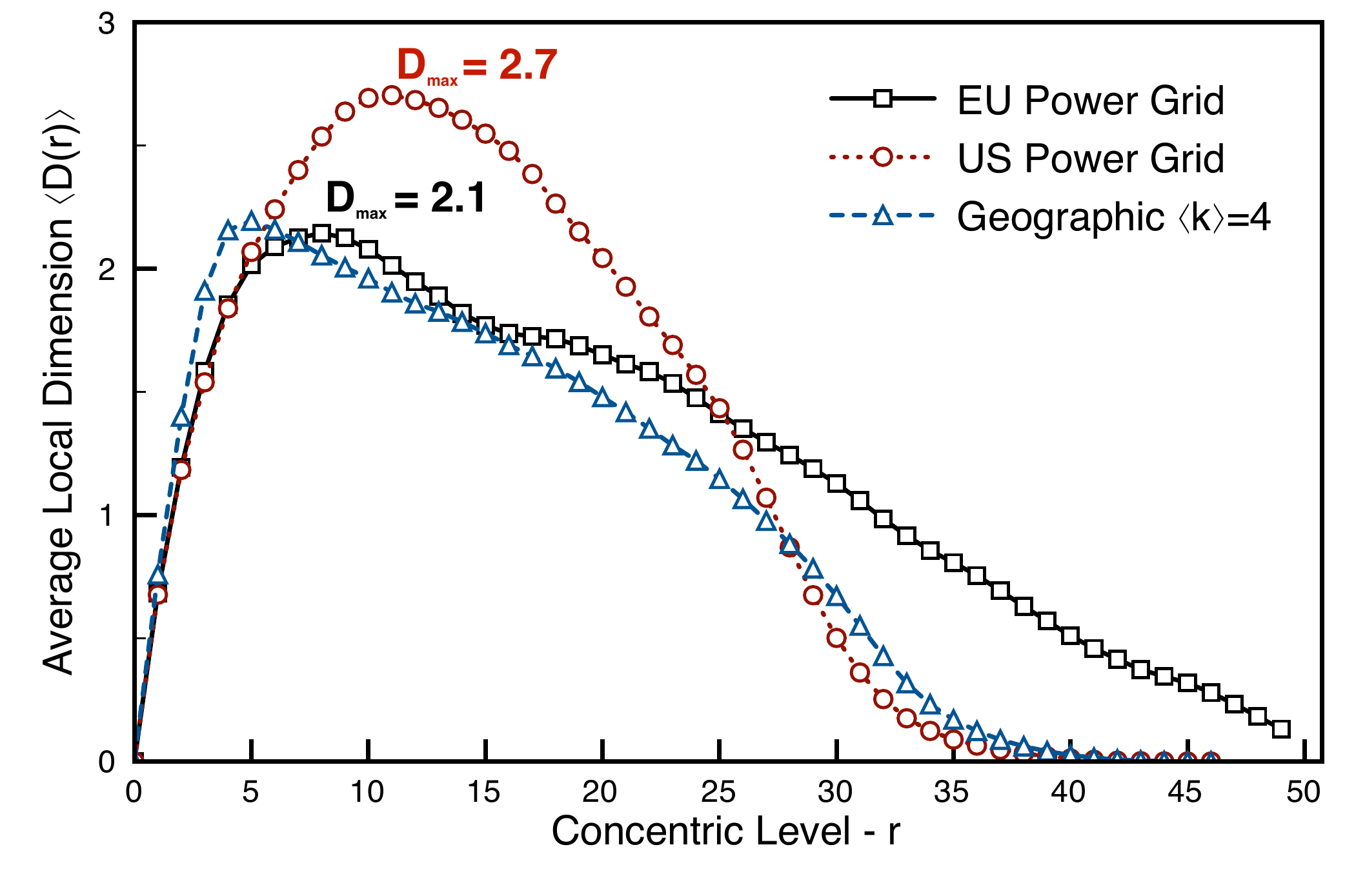}}
 \caption{The EU Network is depicted in ({\bf a}) with colors representing the countries according to the legend. A projection for the structure of the US power grid network is displayed in ({\bf b}), with the node degree represented by a color spectrum. The average local dimension, obtained from equation~\ref{eq:dimensiondiscrete} taken over all nodes of the networks, is shown in  ({\bf c}). The peak of the curves characterizes the maximum dimensionality of such networks. As previously suggested, the embedding space dimension may differ from the topological dimension. For the sake of comparison, the curve for a geographic network model compatible with the power grid network is also shown.}
 \label{fig:powergridProjections}
\end{figure*}

We obtained the local dimensionality patterns for the power grid networks, shown in figure~\ref{fig:averageDimensionVsLevel}. The EU power grid network is characterized by a peak with maximum dimension value near $2$ ($D_{max} = 2.14$ with our method), indicating that the network is strongly planar and may be embedded on a 2D surface without losing most of its topological features.  On the other hand the maximum dimension obtained for the US power grid is $D_{max} =2.7$, suggesting that some networks may present topological dimension different from that of their original embedding space. The global dimensionality values obtained by this methodology are consistent to those obtained previously by Havlin et al\cite{Havlin2011} by directly taking the scaling power exponent.

The curves shown in figure~\ref{fig:averageDimensionVsLevel} also provide important statistical information about the dimensional behavior of diffusion processes and self avoiding random walks along different distances from a central node. While both power grids present a peak of dimensionality, the curve for EU network follows a mostly linear slow decay, differently from the concave (and faster) decay observed for the US network. We see that for very far distances the diffusion takes place on a much more restricted set of nodes, which leads to less degrees of freedom and consequently lower dimension.  This effect occurs much earlier for the US power grid than for the EU network.

Local dimension methodology can also be applied to characterize the nodes themselves, as exemplified in figure~\ref{fig:cityExample}. By embedding a network on a $n$-dimensional space, we can observe the correspondence relation between local topological dimension and the spreading of nodes across each degree of freedom (or axis) available on this space. A two-dimensional lattice or manifold, for example, will present no local positional spreading over a third axis when embedded on the 3D space. In general, if the dimension of the embedded space is higher than the topological dimension, we can say that the network topology is well accommodated in this space.

To illustrate this idea we employed the \emph{Fruchterman-Reingold}\cite{FRUCHTERMAN:1991fk} (FR) force-directed method to embed the power grid networks on the 3D space. This approach allows us not only to understand the spreading of nodes across each axis, but also to effectively visualize the local dimension distribution for the network.

\begin{figure*}[!htbp]
 \centering
 \subfloat[EU Power Grid - front view]{\label{fig:EUPowergrid3DFront}\includegraphics[height=7.75cm]{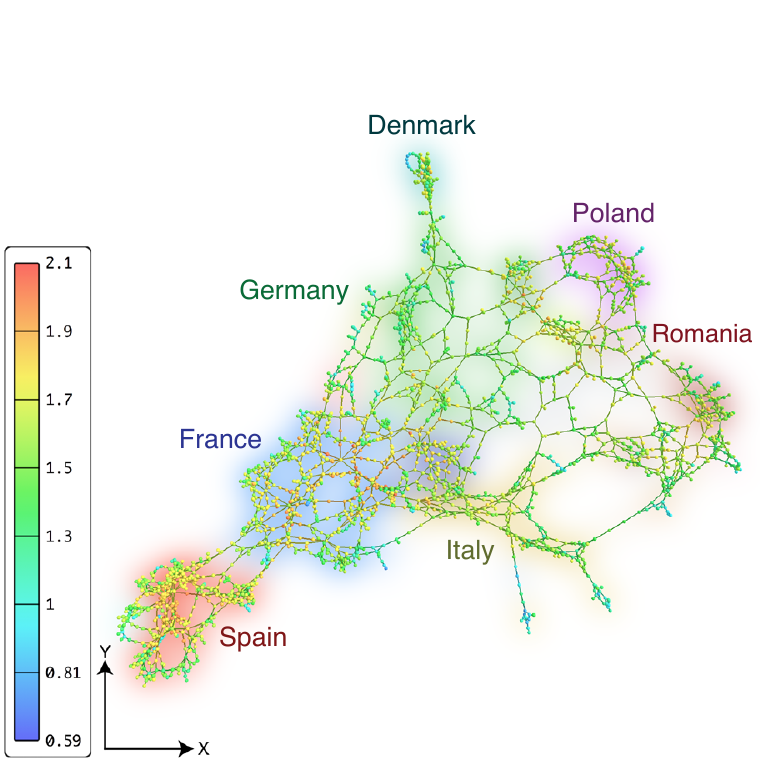}}
 ~
 \subfloat[EU side view]{\label{fig:EUPowergrid3DFront}\includegraphics[height=7.125cm]{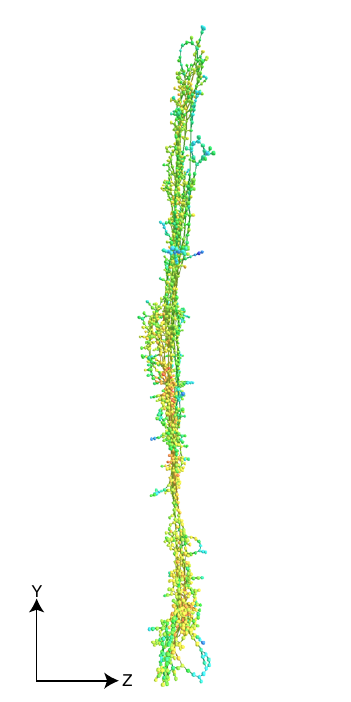}}
 \\
 \subfloat[US Power Grid - front view]{\label{fig:USPowergrid3DFront}\includegraphics[height=7.75cm]{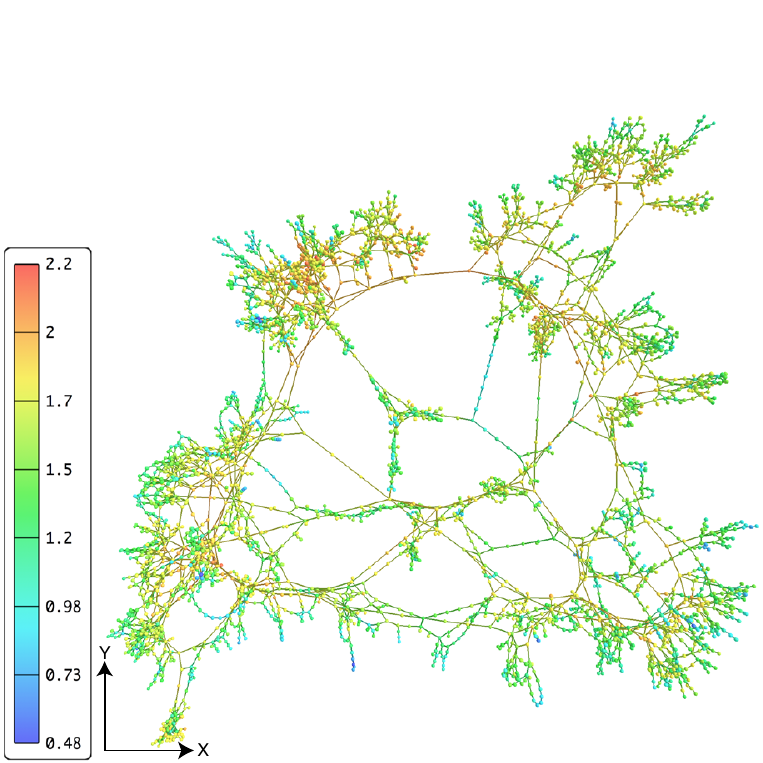}}
 ~
 \subfloat[US side view]{\label{fig:USPowergrid3DFront}\includegraphics[height=7.25cm]{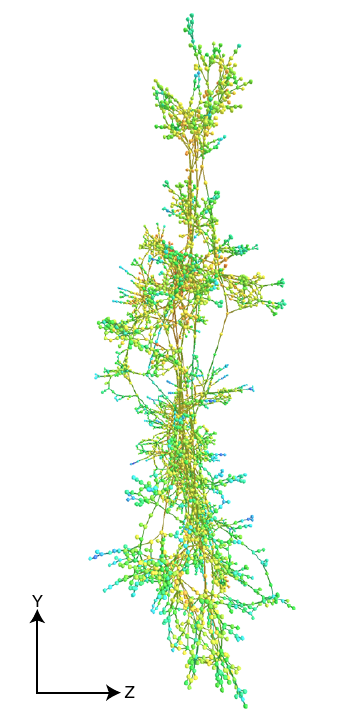}}
 \caption{Power grid networks embedded on the 3D space by using the FR algorithm. Each network is displayed from two distinct viewports: one taken from the two maximum spreading axis, $XY$, depicted in ({\bf a}) and ({\bf c}); and another from its side, $ZY$ in ({\bf b}) and ({\bf c}). Colors indicate the local dimension of nodes at level $r=5$. Countries covered by EU Power Grid are shown as background clouds according with the legend of figure~\ref{fig:EUPowergrid2D}.}
 \label{fig:powergridProjections}
\end{figure*}

Projections of the power grid networks in the 3D space are shown in figure~\ref{fig:powergridProjections}, local dimensions for each node considering the concentric level $r=5$ are represented by colors. As expected, both networks seem to present two-dimensional, geographical nature, with nodes spreading much more on the plane $XY$ than on the axis $Z$. This effect is even more apparent for the EU network, which is more planar than the US network, corroborating the previous results that the dimension of EU network is $2$ while US power grid presents dimension between $2$ and $3$. Nodes with high dimensionality are mostly distributed on dense regions of the networks. On the EU power grid network, Spain and France are characterized by very high dimensionality, while Eastern Europe presents lower dimensionality. This may be the consequence of the fact that most of power plants are located in France and Spain.

Nodes with high local dimension became centers of the spanning of nodes across all three dimensions.  This behavior is more clearly observed in the US power grid  because it is characterized by higher dimensionality. Another observation is that the borders of the networks, in general, present lower dimension when compared with the rest of the network, which is a reflect of what happens in finite regular structures, on a cube, for example.  While its internal topology can be characterized by dimension $3$, its borders are surfaces, mostly characterized by dimension $2$.

\begin{figure*}[!htbp]
 \centering
 \includegraphics[width=12.9cm]{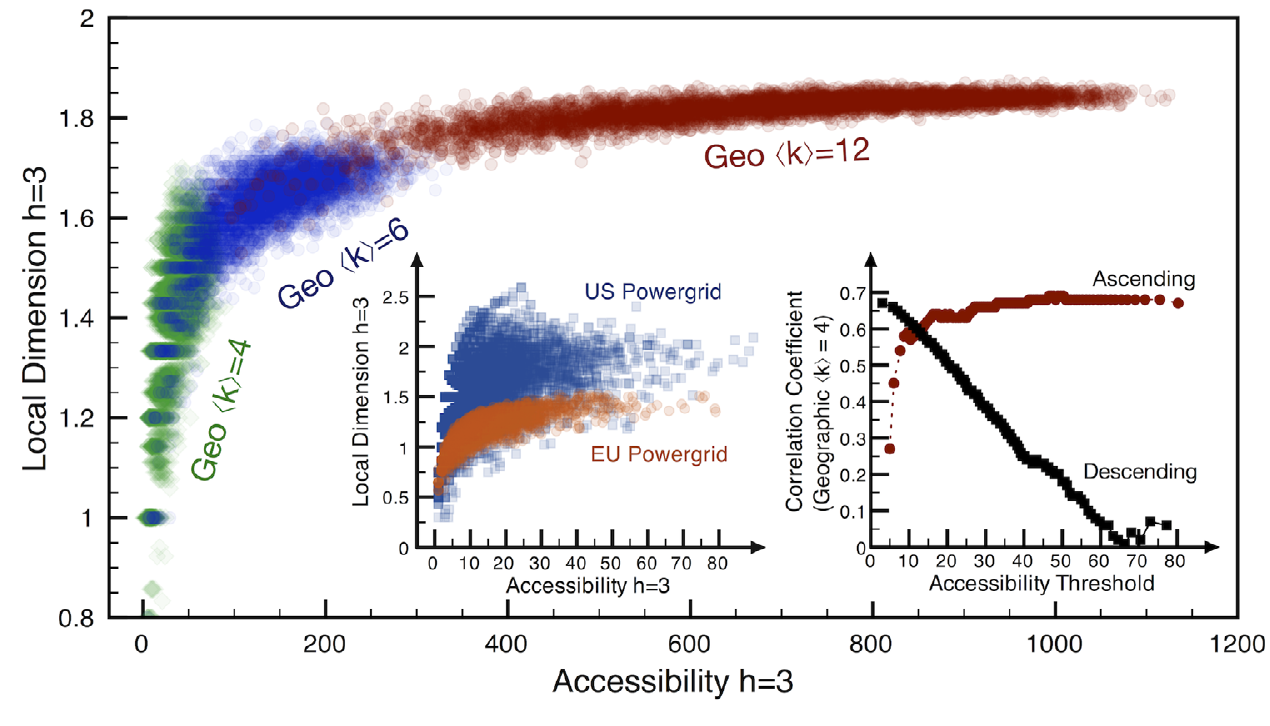}
 \caption{Mapping of the local dimension of nodes against their value of accessibility considering the concentric third concentric level obtained for three generated geographic networks with distinct average degrees and $N=5000$. The left side inset shows the same type of mapping for the US power grid and EU power grid networks while the right side displays the curves of the correlation between both measurements for a geographic network model when only nodes above (ascending) or below (descending) a threshold value of accessibility are considered in the calculations.}
 \label{fig:DimensionVsAccessibility}
\end{figure*}

To further understand the border effect on the local dimensionality we compared it to the node accessibility measurement. Accessibility, $\kappa_r(i)$, can be understood as the effective number of nodes which are accessible in direct pathways (i.e. paths defined by self-avoiding walks) starting from a node $i$ at distance $r$. Small values of accessibility are known to highlight the topological borders of networks much more effectively than by means of other centrality measurements\cite{Travencolo2008Accessibility}. Figure~\ref{fig:DimensionVsAccessibility} presents the plot of accessibility vs local dimension considering $r=3$ for three random geometric networks. A semi-linear relationship between the two measurement can be inferred for each network as nodes with low local dimension tends also to display low accessibility. This result indicates that for these networks the proximity with the borders of the network yields low dimensionality.

Table~\ref{table:AccessibilityCorrelations} shows that for the generated geographical models, local dimension correlates significantly with accessibility. Nodes at the border of these networks contribute much more to the correlation between the two measurements than nodes with higher accessibility, this effect is displayed in the inset on the right side of figure~\ref{fig:DimensionVsAccessibility}, which shows that the correlation rises abruptly to the stable value when the nodes at the border are taken into account first (ascending, displayed in red), than when highly accessible nodes are considered first.

The relationship between accessibility and dimensionality was also obtained for the considered power grid networks(also in table~\ref{table:AccessibilityCorrelations}), with EU power grid presenting high correlation in the same fashion as the geographical models, while the US network presents a much lower value for correlation. The inset on the left side of figure~\ref{fig:DimensionVsAccessibility} makes clear that while EU network displays a mapping very similar to the geographical models, the US power grid network presents a much more sparse and uncorrelated distribution among accessibility and dimensionality with no clear relationship between the two measurements . This may be due to the fact that the US power grid network presents a much more heterogeneous geographical topology in the sense that the geographical role of each node is not well defined, like the other networks. For instance, by observing its 3D projection in figure~\ref{fig:USPowergrid3DFront}, one cannot precisely identify the borders and the center of the network, on the contrary, EU power grid resembles the original shape of the original embedding space (as seen in figure~\ref{fig:EUPowergrid3DFront}), thus providing a natural description of the border region for this network. 

\begin{table}[!htdp]
\caption{Pearson correlation coefficient between accessibility and local dimensionality with $r=3$ for the considered networks.}
\begin{center}
\begin{tabular}{l r}
\hline
Network & Correlation \\ \hline
Geographic $k=4$ & $0.67$ \\
Geographic $k=6$ & $0.68$ \\
Geographic $k=12$ & $0.77$ \\
US Power Grid & $0.50$ \\
EU Power Grid & $0.74$ \\
\hline
\end{tabular}
\end{center}
\label{table:AccessibilityCorrelations}
\end{table}%

We extended the methodology to characterize the dimension of networks and introduced an experimental measurement of local dimensionality of nodes in a network, thus providing more information not only about the global overview of the distribution of dimension along networks, but also about the topology around individual nodes. This method allowed us to identify two distinct regions of dimensionality in coupled regular networks and study the heterogeneous distribution of dimensionality on power grid networks. While the EU power grid displays a mostly planar structure considering both to the topological and embedding space dimensions, the US power grid is clearly less planar. These results were confirmed by applying a force-directed embedding method. Additionally, local dimension measurements were also capable to reveal borders on such networks.

In summary, the combination of the methodologies exploited in this paper provides new insights about the concept of dimensionality in complex networks. The local definition of dimensionality may also provide insights about the local resilience of a network in the same fashion that a 2D lattice or a 1D chain is more susceptive to break apart by being randomly attacked than a 3D block. For a plane one may need to just draw a line to divide it in two, while for a block, it is needed to be cut by a plane. This also can be understood as a result of the fact that higher dimensional structures present higher number of redundant pathways. Further investigations shall be done to understand the consequences of the different dimensional structures with respect to other measurements and dynamics.

L. da F. Costa is grateful to FAPESP (05/00587- 5 and 2011/50761-2) and CNPq (301303/06-1 and 573583/2008-0) for financial support. Filipi N. Silva is grateful to CAPES for sponsorship. The authors also are thankful to the European Network of Transmission System Operators for Electricity (ENTSO-E), for providing supplemental data of the european power grid network.

\bibliography{LocalDimension}

\end{document}